\documentclass[aps,prb,fleqn,12pt,superscriptaddress]{revtex4-1}
\usepackage{epsfig}
\usepackage{graphicx}
\usepackage{pdfpages}
\usepackage{braket}
\usepackage{amssymb,amsmath}
\usepackage{hyperref}
\usepackage{framed}
\usepackage{epstopdf}

\begin{document}

\title{Magnetic excitations in a three-orbital model \\ 
for the strongly spin-orbit coupled iridates: \\ 
Effect of mixing between the \boldmath$J=1/2$ and 3/2 sectors}
\author{Shubhajyoti Mohapatra}
\affiliation{Department of Physics, Indian Institute of Technology Kanpur - 208016, India}
\author{Jeroen van den Brink}
\affiliation{IFW Dresden, Helmholtzstrasse 20, 01069 Dresden, Germany}
\author{Avinash Singh}
\email{avinas@iitk.ac.in}
\affiliation{Department of Physics, Indian Institute of Technology Kanpur - 208016, India}
\affiliation{IFW Dresden, Helmholtzstrasse 20, 01069 Dresden, Germany} 

\date{\today} 
\textwidth 6.3in

\begin{abstract}
A three-orbital-model approach for studying spin wave excitations in the strongly spin-orbit coupled layered perovskite iridates is presented which provides a unified description of magnetic excitations as well as the electronic band structure. The calculated spin wave dispersions with realistic three-band parameters are in excellent agreement with the RIXS data for iridates, including the strong AF zone boundary dispersion in the single-layer compound $\rm Sr_2 Ir O_4$ and the large anisotropy gap in the bilayer compound $\rm Sr_3 Ir_2 O_7$. The RIXS spin wave data is shown to provide evidence of mixing between the $J=1/2$ and 3/2 sectors in both compounds.   


\end{abstract}


\maketitle

\newpage

\section{Introduction}

Magnetic excitations in strongly spin-orbit coupled $5d$ transition metal systems are of considerable recent interest in view of their novel electronic and magnetic properties involving magnetic ordering with spin-orbital entangled states. Resonant inelastic X-ray scattering (RIXS) studies of spin wave excitations in the quasi-two-dimensional antiferromagnet (AF) $\rm Sr_2 Ir O_4$ have shown surprisingly large AF zone boundary dispersion, with spin wave energy at $(\pi/2,\pi/2)$ nearly half of that at $(\pi,0)$.\cite{kim2012magnetic} In the bilayer iridate compound $\rm Sr_3Ir_2O_7$, RIXS studies have revealed distinct magnon dispersion with exceptionally large gap ($\approx$ 80 meV) that is comparable to the magnon bandwidth.\cite{kim2012large,boseggia2015evidence} 

Inelastic neutron scattering experiments on the honeycomb antiferromagnets $\rm A_2 Ir O_3$ (A=Li,Na) have provided overall energy scale of spin excitations and relevant magnetic interactions for the novel magnetic states of composite spin-orbital moments including N\'{e}el, zigzag, and stripy magnetic order predicted in these compounds.\cite{choi2012spin} 
X-ray resonant magnetic scattering and RIXS studies of epitaxially strained $\rm Sr_2 Ir O_4$ thin films have shown extremely anisotropic magnetic correlations, and also tunability of zone-boundary magnon mode energies with compressive/tensile strain.\cite{lupascu2014tuning} Recent RIXS measurements on the pyrochlore osmate $\rm Cd_2 Os_2 O_7$, which exhibits a metal-insulator transition (MIT) closely connected to the magnetic order, have provided insight into the nature of magnetic excitations and the role of SOC in magnetically mediated MIT in $5d$ systems.\cite{calder2016spin} Several of above studies show that substantial further-neighbor exchange couplings are required to explain the observed dispersion.

In particular, $\rm Sr_2IrO_4$ has attracted considerable attention due to its structural and magnetic resemblance with the cuprate antiferromagnet $\rm La_2CuO_4$. Magnon dispersion in $\rm Sr_2IrO_4$ has been theoretically studied in terms of the $J_{\rm eff}=1/2$ quantum Heisenberg antiferromagnet (QHAF) on a square lattice,\cite{kim2012magnetic} rendering the effective low-energy physics similar to that in $\rm La_2CuO_4$. This is further supported by the observation of exciton modes in $\rm Sr_2IrO_4$, whose dispersion is strongly renormalized by magnons, in analogy with hole dynamics in AF background as in cuprate antiferromagnets.\cite{kim2014excitonic} 

Anisotropic spin interactions play an important role in magnetic ordering and excitations in the layered perovskite iridates. In the half-filled $J=1/2$ sector resulting from strong SOC, spin-dependent hopping due to mixing between $yz$ and $xz$ orbitals arising from staggered $\rm Ir O_6$ octahedral rotations lead to pseudo-dipolar (PD) and Dzyaloshinskii-Moriya (DM) interactions in addition to the isotropic Heisenberg  interaction,\cite{jackeli2009mott,jin2009,carter2013theory} which are order-of-magnitude larger than in 3d transition metal compounds.\cite{katukuri14basal} The inherent frustration in the bilayer system due to different canting proclivities of in-plane and out-of-plane neighboring spins breaks the degeneracy between c-axis orientation and planar canted configuration (as in the monolayer system), resulting in c-axis AF ordering and strong anisotropy-induced spin-wave gap in the bilayer iridate.

In addition to magnetic excitations in the insulating state, RIXS studies of doping evolution of AF order and comparison with hole doped cuprates is also of strong current interest.\cite{gretarsson2016doping,liu2016anisotropic} Unconventional superconductivity has been predicted in doped $\rm Sr_2IrO_4$,\cite{yang2014superconductivity,yan2015signature,gao2015possible} evolution from long-range to short-range AF order with doping has been demonstrated in single layer and bilayer strontium 
iridates,\cite{chen2015influence,gretarsson2016doping} and various exotic properties such as Fermi arcs, pseudogaps in electron doped $\rm Sr_2IrO_4$, doping evolution of the strong interlayer coupling, and doping driven insulator-to-metal transition in $\rm Sr_3Ir_2O_7$ have been discovered recently.\cite{kim2014fermi,he2015fermi,kim2016observation,lu2016electron} 

Most theoretical investigations of magnetic excitation spectra have been carried out within localized spin models.\cite{kim2012magnetic, kim2012large,boseggia2015evidence,gretarsson2016doping,lu2016electron} Spin-wave spectrum in $\rm Sr_2IrO_4$ has been described in terms of phenomenological $J$-$J^{\prime}$-$J^{\prime\prime}$ Heisenberg model, where the strong dispersion along AF zone boundary $(\pi,0) \rightarrow (\pi/2,\pi/2) \rightarrow (0,\pi)$ was ascribed to ferromagnetic $J^\prime$ coupling.\cite{kim2012magnetic} 
This is as expected from earlier studies of magnon spectrum in the Hubbard model, where it was shown that effective ferromagnetic $J^\prime$ spin coupling is generated in the intermediate coupling region, which accounts for the measured zone boundary magnon dispersion in the cuprate antiferromagnet $\rm La_2 Cu O_4$.\cite{singh1993spin,singh2002spin} Effects of finite SOC, finite $U$, and finite orbital mixing arising from $\rm IrO_6$ octahedral rotation have been considered mostly in the context of electronic band structure. 


In this paper, we will therefore consider an approach in which spin dynamics can be directly studied within the three-orbital model, which will allow all three effects mentioned above to be incorporated simultaneously on equal footing. Based on a realistic three-orbital model involving $xz$, $yz$, and $xy$ Ir $5d$ orbitals, this approach can provide a unified description of electronic band structure as well as magnetic excitations in both single-layer and bilayer compounds. In particular, we will investigate whether the RIXS spin wave spectra for both single-layer and bilayer iridates provide evidence of mixing between the $J=1/2$ and 3/2 sectors. The itinerant-electron approach for studying magnetic excitations can also be naturally extended to finite doping studies which are of strong current interest. The intermediate coupling region in between Slater and Mott type insulators has been suggested as appropriate for the moderately correlated iridate compounds.\cite{watanabe2014theoretical}
Although RIXS in $\rm Sr_2 Ir O_4$ has been analyzed in terms of a multi-orbital itinerant electron 
approach where magnetic excitations were studied in the Hartree-Fock and random phase approximations,\cite{igarashi_prb_2014} realistic band structure and application to the bilayer compound have not been 
considered.

\section{Three-orbital model and electronic band structure}
The distinctive electronic and magnetic properties in the layered Ruddelsden-Popper series of iridates are determined by the Ir$^{4+}$ ions in 5$d^5$ electronic configuration, sitting in the octahedral crystal field of oxygen. The strong crystal field splits the 5$d$ levels into two e$_g$ and three t$_{2g}$ levels. Spin-orbit coupling (SOC) further splits the t$_{2g}$ multiplets into (upper) $J=1/2$ doublet and (lower) $J=3/2$ quartet with an energy gap of $3\lambda/2$. Four of the five electrons fill the $J=3/2$ states, leaving one electron for the $J=1/2$ sector, rendering it magnetically active in the ground state.\cite{kim2008novel,kim2009phase} Due to the large energy separation ($\sim$3 eV) of e$_g$ levels, low energy physics is effectively described by projecting out the empty e$_g$ levels. In the t$_{2g}$ manifold, the states $|J,m \rangle$ have the form:
\begin{eqnarray}
\Ket{\frac{1}{2},\pm\frac{1}{2}} &=& \left [\Ket{yz,\bar{\sigma}} \pm i \Ket{xz,\bar{\sigma}} \pm \Ket{xy,\sigma}\right ] / \sqrt{3} \nonumber \\
\Ket{\frac{3}{2},\pm\frac{3}{2}} &=& \left [\Ket{yz,\sigma} \pm i \Ket{xz,\sigma}\right ] / \sqrt{2} \nonumber \\
\Ket{\frac{3}{2},\pm\frac{1}{2}} &=& \left [\Ket{yz,\bar{\sigma}} \pm i \Ket{xz,\bar{\sigma}} \mp 2 \Ket{xy,\sigma} \right ] / \sqrt{6} 
\label{jmbasis}
\end{eqnarray}
where $\Ket{yz,\sigma}$, $\Ket{xz,\sigma}$, $\Ket{xy,\sigma}$ are the t$_{2g}$ states and the signs $\pm$ correspond to spins $\sigma = \uparrow/\downarrow$. The coherent superposition of different-symmetry orbitals, with opposite spin polarization between $xz$/$yz$ and $xy$ levels implies spin-orbital entanglement, and also imparts unique extended 3D shape to these states, which has important consequence in the bilayer compound, as discussed in Sec. IV. 

\subsection{Non-magnetic state}
We will consider a three-orbital model within the t$_{2g}$ manifold which allows for investigation of role of mixing between the $J=1/2$ and 3/2 sectors on magnetic excitations in both single-layer and bilayer iridate compounds. We start with the free part of the Hamiltonian including the local spin-orbit coupling and the band terms represented in the three-orbital basis $(yz\sigma,xz\sigma,xy\bar{\sigma})$: 
\begin{equation}
\mathcal{H}_{\rm SO} + \mathcal{H}_{\rm band} = \sum_{{\bf k} \sigma} \psi_{{\bf k} \sigma}^{\dagger} \begin{pmatrix}
{\cal E}_{\bf k} ^{yz} & i \sigma\frac{\lambda}{2} & -\sigma\frac{\lambda}{2} \\
- i \sigma\frac{\lambda}{2} & {\cal E}_{\bf k} ^{xz} & i\frac{\lambda}{2} \\
-\sigma\frac{\lambda}{2} & - i\frac{\lambda}{2} & {\cal E}_{\bf k} ^{xy} \\
\end{pmatrix} \psi_{{\bf k} \sigma} 
\label{three_orb_matrix}
\end{equation} 
where ${\cal E}_{\bf k} ^\mu$ ($\mu=yz,xz,xy$) are the band energies for the three orbitals. In the following it will be convenient to distinguish between the band energy contributions from hopping terms connecting opposite sublattices ($\epsilon_{\bf k} ^{\mu}$) and same sublattice (${\epsilon_{\bf k} ^{\mu}}^\prime$). It is important to note that effective electron hopping occurs between spin-orbital entangled states on neighboring sites. 

\subsection{AF state and staggered field}
Including the ($z$-direction) symmetry-breaking staggered field $-s\tau\Delta$ in the $J=1/2$ sector (where $s=\pm 1$ for the two sublattices A/B, and $\tau =\pm 1$ for the two pseudospins $\uparrow / \downarrow$), and transforming back to the three-orbital basis, the staggered-field contribution is obtained as: 
\begin{equation}
\mathcal{H}_{\rm sf} = \sum_{{\bf k} \sigma s} 
\frac{s \sigma \Delta}{3} 
\psi_{{\bf k} \sigma s}^{\dagger} 
\begin{pmatrix}
1 & i \sigma & -\sigma \\
-i \sigma & 1 & i \\
-\sigma & -i & 1 \\
\end{pmatrix} \psi_{{\bf k} \sigma s} 
\label{stag_field_matrix}
\end{equation} 
which has exactly same structure as the spin-orbit coupling term. The above transformation from $J=1/2$ basis to three-orbital basis is illustrated below. Starting with the local field term for $\sigma=\uparrow$ (pseudospin $\tau = \downarrow$ in the $J=1/2$ basis), we obtain:
\begin{eqnarray}
\mathcal{H}_{\rm sf} ^{\sigma=\uparrow} &=& \Delta |1/2,-1/2 \rangle \langle 1/2,-1/2| \nonumber \\
& = & \frac{\Delta}{3} (|yz \uparrow \rangle - i|xz\uparrow \rangle -|xy\downarrow \rangle ) (\langle yz \uparrow | + i \langle xz\uparrow | -\langle xy\downarrow | ) \nonumber \\
& = & \frac{\Delta}{3} 
\begin{pmatrix} 
| yz \uparrow \rangle & | xz\uparrow \rangle & | xy\downarrow \rangle 
\end{pmatrix} 
\begin{pmatrix}
1 & i & -1 \\
-i & 1 & i \\
-1 & -i & 1 \\
\end{pmatrix} \begin{pmatrix} 
\langle yz \uparrow | \\ \langle xz\uparrow | \\ \langle xy\downarrow | 
\end{pmatrix} 
\end{eqnarray}   
in the three-orbital basis $(|yz \uparrow \rangle,|xz\uparrow \rangle,|xy\downarrow \rangle )$. 

Combining all three terms above, the total Hamiltonian is given below in the composite three-orbital, two-sublattice basis, showing the hopping terms connecting same and opposite sublattices, and the staggered fields on the two sublattices. Also included are the hopping terms involving orbital mixing between $yz$ and $xz$ orbitals due to the octahedral rotation. Involving nearest-neighbor hopping, these orbital mixing terms are placed in the sublattice-off-diagonal $(s\bar{s})$ part of the Hamiltonian:
\begin{eqnarray}
& & \mathcal{H}_{\rm SO} + \mathcal{H}_{\rm band} + \mathcal{H}_{\rm sf} 
 \nonumber \\ & = & \sum_{{\bf k} \sigma s} \psi_{{\bf k} \sigma s}^{\dagger} \left [ \begin{pmatrix}
{\epsilon_{\bf k} ^{yz}}^\prime & i \sigma\frac{\lambda}{2} + {\epsilon^{yz|xz}_{k}}^{\prime} & -\sigma\frac{\lambda}{2} \\
- i \sigma\frac{\lambda}{2} + {\epsilon^{yz|xz}_{k}}^{\prime} & {\epsilon_{\bf k} ^{xz}}^\prime & i\frac{\lambda}{2} \\
-\sigma\frac{\lambda}{2} & - i\frac{\lambda}{2} & {\epsilon_{\bf k} ^{xy}}^\prime \end{pmatrix} 
+ \frac{s \sigma \Delta}{3}
\begin{pmatrix}
1 & i \sigma & -\sigma \\
-i \sigma & 1 & i \\
-\sigma & -i & 1 \\
\end{pmatrix} \right ]
\psi_{{\bf k} \sigma s} \nonumber \\
& + & 
\sum_{{\bf k} \sigma s} \psi_{{\bf k} \sigma s}^{\dagger}
\begin{pmatrix}
\epsilon_{\bf k} ^{yz} & \epsilon_{\bf k} ^{yz|xz} & 0 \\
-\epsilon_{\bf k} ^{yz|xz} & \epsilon_{\bf k} ^{xz} & 0 \\
0 & 0 & \epsilon_{\bf k} ^{xy} \end{pmatrix} 
\psi_{{\bf k} \sigma \bar{s}} 
\label{three_orb_two_sub}
\end{eqnarray} 
The self-consistent determination of the staggered field $\Delta$ in terms of the densities $\langle n_{i\tau}  \rangle$ for the two pseudospins $\tau = \uparrow,\downarrow$ of the $J=1/2$ sector is discussed in Sec. III. 

For straight Ir-O-Ir bonds, all hopping terms between NN Ir ions are orbital-diagonal with no orbital mixing. Due to twisting of Ir-O-Ir bonds associated with rotation of $\rm IrO_6$ octahedra in $\rm Sr_2IrO_4$ and $\rm Sr_3Ir_2O_7$, local cubic axis of $\rm IrO_6$ octahedra are alternatively rotated, giving rise to significant mixing between the $yz$ and $xz$ orbitals. In the $J=1/2$ sector, the NN orbital mixing yields spin-dependent hopping terms (Appendix B) which break spin rotation symmetry. However, in $\rm Sr_2IrO_4$, these spin-dependent hopping terms can be gauged away by a spin- and site-dependent  unitary transformation (Appendix C), leaving the usual SU(2)-invariant model in the $J=1/2$ sector.\cite{senthil2011twisted} 

In the bilayer compound $\rm Sr_3Ir_2O_7$, on the other hand, due to different magnitude of spin-dependent hopping terms between in-plane and out-of-plane neighbors, both cannot be gauged away by the same transformation. As explained in Appendix D, this results in a magnetic frustration which accounts for the $c$-axis spin orientation and the large anisotropy gap seen in the RIXS spin wave spectrum.  


\subsection{Tight-binding model for $\rm Sr_2IrO_4$} 
This is a layered compound with unit cell containing four $\rm IrO_2$ layers. We will focus on a single layer and neglect inter-layer hopping. Due to staggered rotation of $\rm IrO_6$ octahedra with respect to c-axis by an angle of $11^\circ$, Ir atoms form two sublattices, leading to doubling of the unit cell,\cite{watanabe2010microscopic,senthil2011twisted} which is in conformity with the staggered magnetic order.  Corresponding to the hopping terms in the tight-binding representation, the various band dispersion terms in Eq. (\ref{three_orb_two_sub}) are given by: 
\begin{eqnarray}
\epsilon^{xy}_{k} &=& -2t_1(\cos{k_x} + \cos{k_y}) \nonumber \\
{\epsilon^{xy}_{k}}^{\prime} &=& - 4t_2\cos{k_x}\cos{k_y} - \> 2t_3(\cos{2{k_x}} + \cos{2{k_y}}) + \mu_{xy}  \nonumber \\
\epsilon^{yz}_{k} &=& -2t_5\cos{k_x} -2t_4 \cos{k_y} \nonumber \\
\epsilon^{xz}_{k} &=& -2t_4\cos{k_x} -2t_5 \cos{k_y} \nonumber \\
\epsilon^{yz|xz}_{k} &=&  -2t_{m}(\cos{k_x} + \cos{k_y}) \nonumber \\
{\epsilon^{yz|xz}_{k}}^{\prime} &=&  -4t^\prime_{m}\sin{k_x}\sin{k_y} \nonumber
\end{eqnarray}
Here $t_1$, $t_2$, $t_3$ are respectively the first, second, and third neighbor hopping terms for the $xy$ orbital, which has energy offset $\mu_{xy}$ from the degenerate $yz/xz$ orbitals induced by the tetragonal splitting. For the $yz$ ($xz$) orbital, $t_4$ and $t_5$ are the NN hopping terms in $y$ $(x)$ and $x$ $(y)$ directions, respectively. Mixing between $xz$ and $yz$ orbitals is represented by the hopping terms $t_m$ (NN) and $t^\prime_{m}$ (NNN). We have taken values of the tight-binding parameters ($t_1$, $t_2$, $t_3$, $t_4$, $t_5$, $t_m$, $t_m^{\prime}$, $\mu_{xy}$) = (0.36, 0.18, 0.09, 0.37, 0.06, 0.0, 0.0, -0.36) eV as considered earlier.\cite{watanabe2010microscopic} In the locally rotated  co-ordinate frame, the orbital mixing terms vanish, and we have therefore set them to zero. 
\begin{figure}
\vspace*{0mm}
\hspace*{0mm}
\psfig{figure=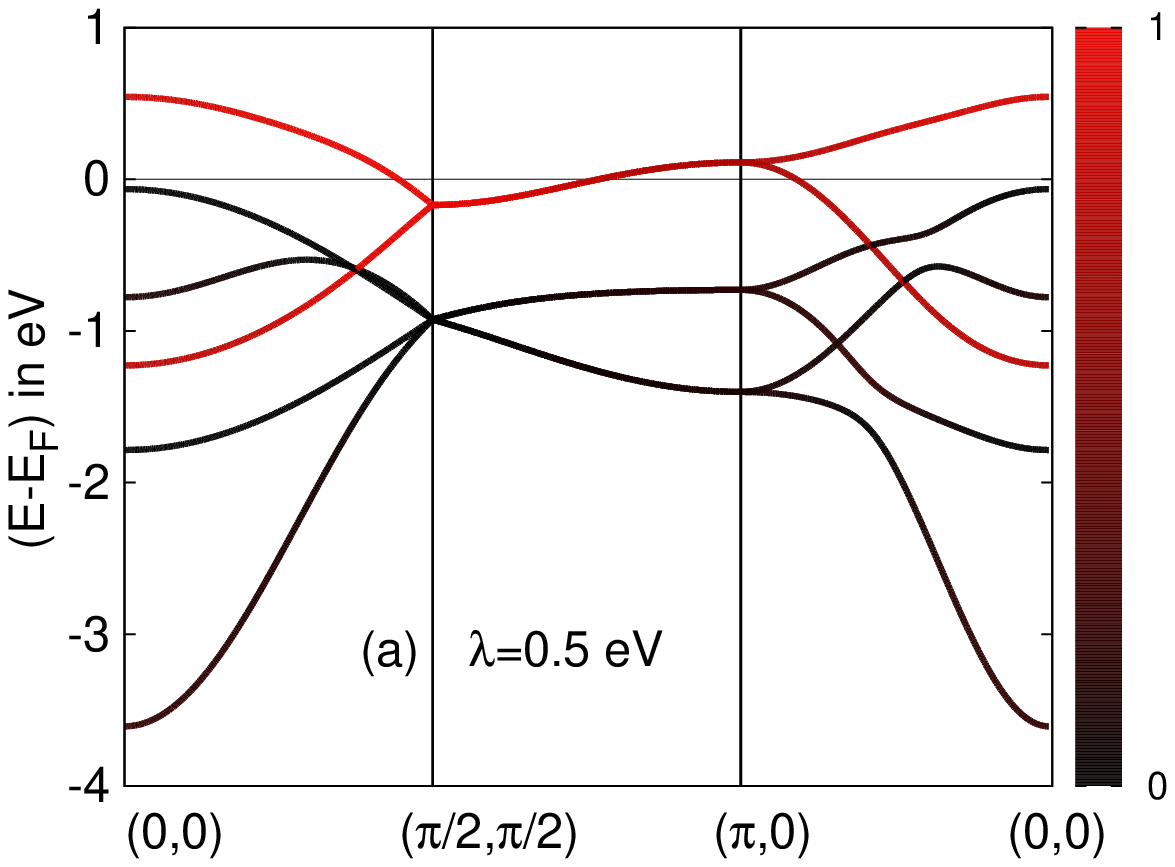,angle=0,width=80mm}
\psfig{figure=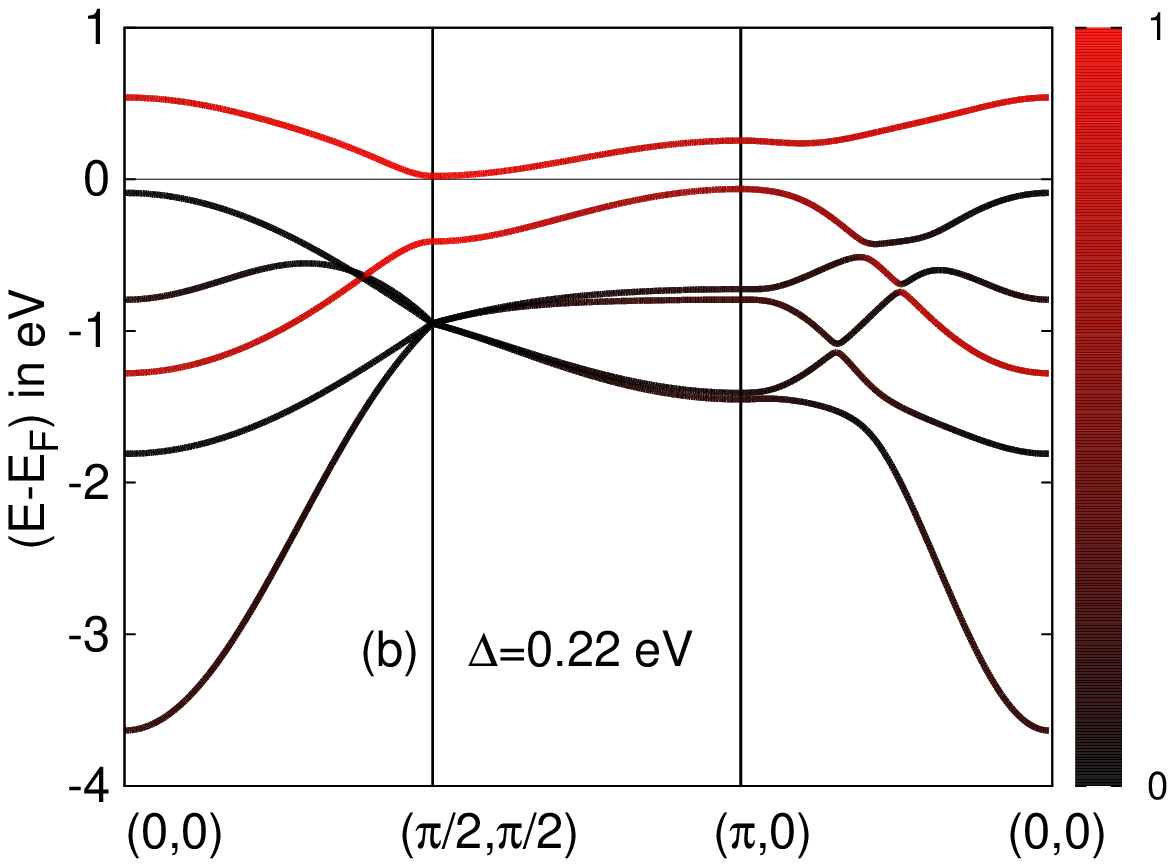,angle=0,width=80mm}
\caption{Electronic band structure for $\rm Sr_2IrO_4$ with SOC only (a), and with both SOC and staggered field (b). The weight of $J=1/2$ (3/2) states are indicated by red (black). The bands are degenerate along the AF zone boundary (a) with the top band crossing the Fermi energy. The staggered field splits the top ($J=1/2$) band (b) resulting in the AF insulating state.} 
\label{mono_bands}
\end{figure}

Figure \ref{mono_bands} shows the band structure for $\rm Sr_2IrO_4$ at half filling with SOC only (a), and with SOC and staggered field (b). In units of $t_1$, we have $\lambda=1.4$ and $\Delta=0.6$. The weightage of states in the $J = 1/2$ and 3/2 sectors are shown as red and black, respectively. The $J=1/2$ weight is calculated using the amplitude-square (summed over two sublattices) of the ${\bf k}$-state eigenvector in the $J=1/2$ basis for a given pseudospin. Although there is significant mixing between $J=1/2$ and $J=3/2$ sectors below the Fermi level, states near Fermi level have mostly $J=1/2$ character at intermediate SOC. The band structure shows distinct Fermi-surface crossing of $J=1/2$ band along the AF zone boundary (a). The bottom (top) of the unoccupied (occupied) band barely touches the Fermi level (b). Topology of bands near Fermi level are significantly affected by the staggered magnetic order for $\rm Sr_2IrO_4$, but are only weakly affected for $\rm Sr_3Ir_2O_7$, as discussed in Sec. IV. At intermediate SOC, the calculated band dispersions as shown above are consistent with first-principles calculations, optical spectroscopy studies, and ARPES data.\cite{kim2008novel,moon2008dimensionality,wang2013dimensionality}

\section{Magnetic excitations in the three-orbital model}
In this section, we introduce the formalism for investigating magnetic excitations in the three-orbital model and apply it to the single layer compound Sr$_2$IrO$_4$. This approach accounts for the mixing between $J=1/2$ and 3/2 sectors, and we will see evidence of this from comparison of the calculated spin wave spectrum with the RIXS data. As the magnetic symmetry breaking is driven by the Hubbard interaction $\sum_i U n_{i\tau} n_{i\bar{\tau}}$ in the $J=1/2$ sector, we will consider the corresponding transverse spin-fluctuation propagator: 
\begin{equation}
\chi^{-+}({\bf q},\omega) = \int dt \sum_{i} e^{i\omega(t-t^\prime)}
e^{-i{\bf q}.({\bf r}_i - {\bf r}_j)}  
\langle \Psi_0 | T [ J_i ^- (t) J_j ^+ (t^\prime) ] | \Psi_0 \rangle
\end{equation}
in terms of the lowering and raising operators $J_{i}^{-}$ and $J_{j}^{+}$ at sites $i$ and $j$. 

In the random phase approximation (RPA), the spin wave propagator is obtained as:
\begin{eqnarray}
[\chi^{-+}({\bf q},\omega)] = \frac{[\chi^0({\bf q},\omega)]}
{1 - U [\chi^0({\bf q},\omega)]}
\label{eq:spin_prop}  \end{eqnarray}
where the bare particle-hole propagator:
\begin{eqnarray}
[\chi^{0}({\bf q},\omega)]_{ss^\prime} & = & i \int \frac{d\omega^{\prime}}{2\pi} \sum_{{\bf k}^{\prime}}  
\left [ G^0 _\uparrow ({\bf k}^{\prime},\omega^{\prime}) \right ]_{ss^\prime} .  \left [ G^0_\downarrow ({\bf k}^{\prime} - {\bf q},\omega^{\prime} - \omega) \right ]_{s^{\prime}s} \nonumber \\ 
& = & \sum_{{\bf k}^\prime,m,n} \left[ 
\frac {
\varphi^s_{{\bf k}^\prime \uparrow m} 
\varphi^{s^{\prime}\ast}_{{\bf k}^\prime \uparrow m} 
\varphi^{s^{\prime}}_{{\bf k^\prime -q}\downarrow n} 
\varphi^{s \ast}_{{\bf k^\prime -q}\downarrow n} } 
{ E^{+}_{{\bf k^\prime -q}\downarrow n} - E^{-}_{{\bf k}^\prime \uparrow m} + \omega -i\eta}  
+ \> \frac{\varphi^{s}_{{\bf k}^\prime \uparrow m} \varphi^{s^{\prime}\ast}_{{\bf k}^\prime \uparrow m} \varphi^{s^{\prime}}_{{\bf k^\prime -q}\downarrow n} \varphi^{s \ast}_{{\bf k^\prime -q}\downarrow n}}{E^{+}_{{\bf k}^\prime \uparrow m} - E^{-}_{{\bf k^\prime -q}\downarrow n} - \omega - i\eta} \right ] 
\end{eqnarray}
is obtained from the Hartree-Fock level Green's functions in the two-sublattice basis by integrating out the fermions in the $(\pi,\pi)$ ordered state. Here $E_{{\bf k}\tau m}$ are the band energy eigenvalues and $\varphi_{{\bf k}\tau m}$ are the wave functions derived by projecting the ${\bf k}$ states in the three-orbital basis onto the basis states $| 1/2,\tau = \pm 1/2 \rangle$ of the $J=1/2$ sector corresponding to pseudospins $\tau = \uparrow,\downarrow$:
\begin{equation}
\varphi_{{\bf k}\uparrow m} = \frac{1}{\sqrt{3}} \left( \phi^{yz}_{{\bf k}\downarrow m} - i\phi^{xz}_{{\bf k}\downarrow m} + \phi^{xy}_{{\bf k}\uparrow m}\right) \;\;\;\;\;\; \varphi_{{\bf k}\downarrow m} = \frac{1}{\sqrt{3}} \left( \phi^{yz}_{{\bf k}\uparrow m} 
+ i \phi^{xz}_{{\bf k}\uparrow m} - \phi^{xy}_{{\bf k}\downarrow m}\right)
\end{equation}
in terms of the wave functions $\phi^{\mu}_{{\bf k}\sigma m}$ in the three-orbital basis $(\mu = yz,xz,xy)$. Here the subscripts $m,n$ indicate eigenvalue branches, $s,s^{\prime}=A/B$ denote sublattice indices, and the superscript +(-) refers to particle (hole) energies above (below) the Fermi energy. Note that the bare particle-hole propagator and therefore the spin wave propagator involve the spin-orbital entangled states. 

From the interaction term $\sum_i U n_{i\tau} n_{i\bar{\tau}}$ in the $J=1/2$ sector, the self-consistency condition is given by $2\Delta = m(\Delta)U$, where the staggered (sublattice) magnetization: 
\begin{equation}
m(\Delta) = [n_A^\uparrow - n_A^\downarrow](\Delta) = [n_A^\uparrow - n_B^\uparrow](\Delta) = \sum_{{\bf k},m} ^{E_{{\bf k}\tau m} < E_{\rm F}}
[\varphi_{{\bf k}\uparrow m} ^2 - \varphi_{{\bf k}\downarrow m} ^2]_A (\Delta).
\end{equation}
In practice, it is easier to consider a given $\Delta$, and self-consistently determine the interaction strength $U=2\Delta/m(\Delta)$, which ensures gapless Goldstone mode in accordance with the spin rotation symmetry in $J=1/2$ space. The $[\chi^{0}({\bf q},\omega)]$ matrix $(2\times 2)$ is evaluated by performing the ${\bf k}$ sum over the two-dimensional Brillouin zone (divided into $300 \times 300$ mesh). In terms of the larger eigenvalue $\lambda_{\bf q}(\omega)$ of the $[\chi^{0}({\bf q},\omega)]$ matrix, the spin wave energies are then evaluated from the pole condition $1-U\lambda_{\bf q}(\omega)=0$ corresponding to Eq. (\ref{eq:spin_prop}).  

\begin{figure}
\vspace*{0mm}
\hspace*{0mm}
\psfig{figure=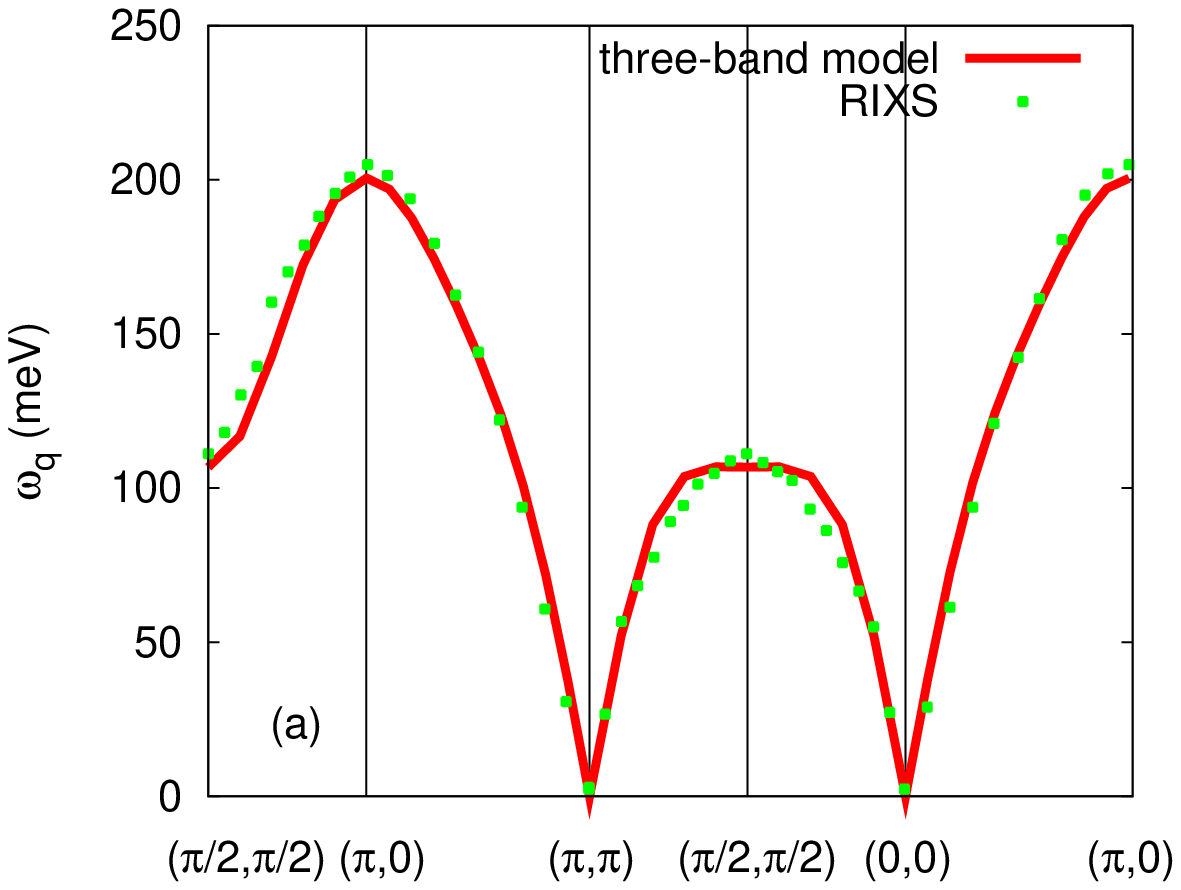,angle=0,width=80mm}
\psfig{figure=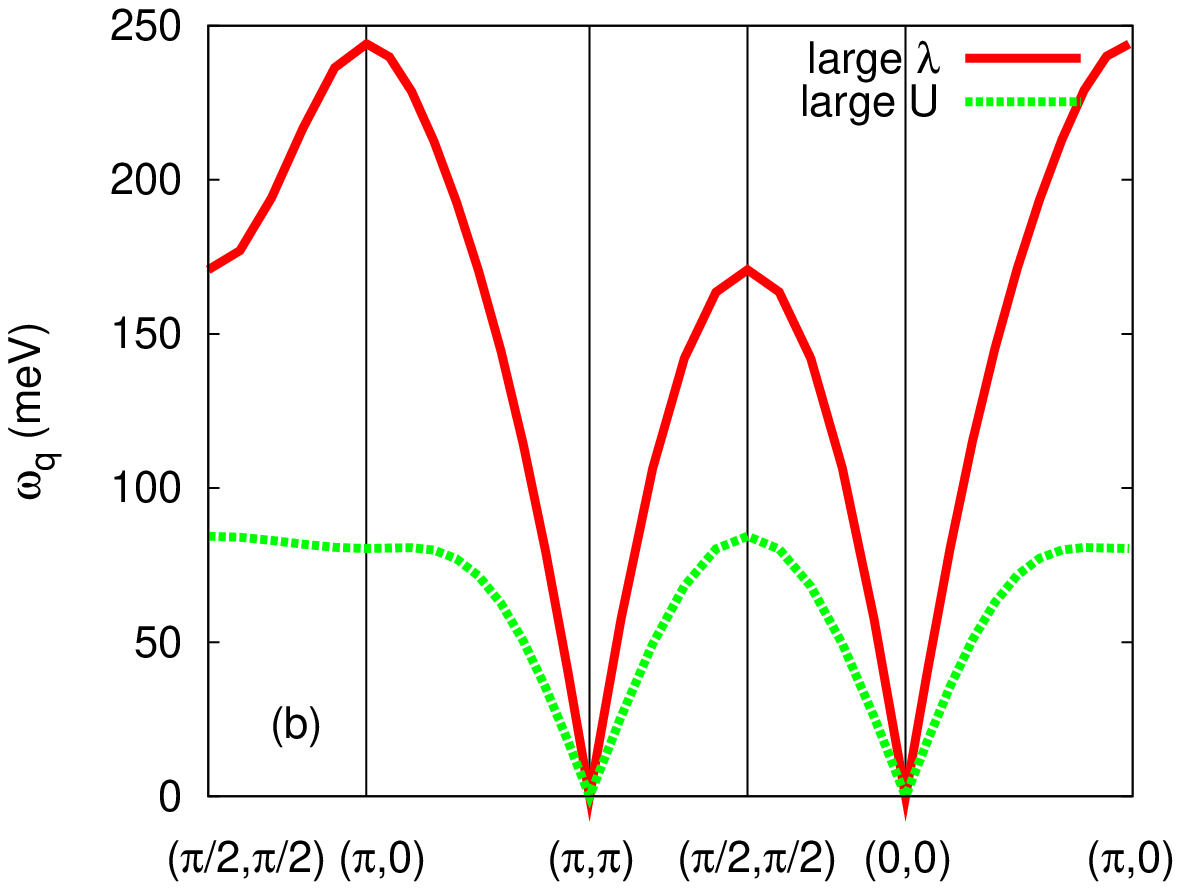,angle=0,width=80mm}
\caption{(a) Calculated spin wave dispersion using the three-orbital-model approach and comparison with RIXS data for $\rm Sr_2 Ir O_4$. (b) In the two limiting cases, the dispersion along the AF zone boundary (from $(\pi/2,\pi/2)$ to $(\pi,0)$) is significantly reduced in comparison to the RIXS data, highlighting the effect of finite mixing between $J=1/2$ and 3/2 sectors and stronger finite-$U$ effect compared to cuprates.} 
\label{fig2}
\end{figure}

The calculated spin wave dispersion using the formalism discussed above is shown in Fig. \ref{fig2}. For the parameter values: $t_1=1$ (energy scale unit = 275 meV), $t_2=0.5$, $t_3=0.25$, $t_4=1.028$, $t_5=0.167$, $\Delta=0.9$, $\lambda=1.35$, $\mu_{xy}=-0.7$, which are nearly same as considered in band structure studies of $\rm Sr_2 Ir O_4$,\cite{watanabe2010microscopic} the calculated spin wave dispersion is in good agreement with the RIXS data. A small fourth neighbor hopping term $(0.2)$ for the $xy$ orbital was included to tune the shape near $(\pi/2,\pi/2)$. Here $U/t_1=2.8$ and $m=0.64$, which correspond to the intermediate-coupling range. The values $\lambda \approx 0.4$ eV and $U \approx 0.8$ eV taken above lie within the estimated energy range for iridates.\cite{kim2008novel,jin2009} The orbital resolved staggered magnetizations and electronic densities are obtained as: $m_{yz}=m_{xz}=-0.22$ and $m_{xy}=0.18$; $n_{yz}=n_{xz}=1.62$ and $n_{xy}=1.76$. 


As the $J=1/2$ and $J=3/2$ atomic states are separated by energy $3\lambda/2$, mixing between the two sectors decreases with $\lambda$, and the two sectors get decoupled at large $\lambda$. One consequence of mixing is that the staggered field in $J=1/2$ sector magnetizes $J=3/2$ states. The magnetic moments developed are, however, smaller by nearly two orders of magnitude due to $J=3/2$ bands being completely filled. A more dramatic effect of finite mixing between the $J=1/2$ and 3/2 sectors is seen in the spin wave dispersion, as discussed below.  

The calculated spin wave dispersions are also shown for comparison [Fig. \ref{fig2}(b)] in two limiting cases: i) large SOC ($\lambda = 20$) and ii) large correlation term ($U \approx 10$), corresponding to negligible mixing between $J=1/2$ and 3/2 sectors and the two sublattices, respectively. All other parameters are same as in Fig. \ref{fig2}(a). We will focus on the ratio $\omega_{(\pi/2,\pi/2)} / \omega_{(\pi,0)}$ of spin wave energies, which provides a quantitative measure of the dispersion along AF zone boundary. Fig. \ref{fig2}(b) shows a significant reduction in the AF zone boundary dispersion at large SOC, with the ratio changing from 0.5 to 0.7. The close agreement of RIXS data with finite SOC calculation provides experimental evidence of mixing between the $J=1/2$ and 3/2 sectors. Similarly, at large $U$, the ratio is nearly 1, highlighting the importance of finite-$U$ effects on spin wave dispersion, as discussed in Appendix A. 

\section{Band Structure and Magnetic Excitations in the Bilayer Compound}


The compound $\rm Sr_3Ir_2O_7$ consists of strongly coupled $\rm Ir O_2$ bilayers stacked along the $c$-axis, with neighboring oxygen octahedra (both intra and interlayer) rotated opposite to each other around $c$-axis by $12^\circ$ angle. The interlayer Ir-O-Ir bonds are straight whereas the intralayer bonds are twisted.\cite{fujiyama2012weak} 
In this section we will extend the three-orbital-model formalism discussed in the previous section to study magnetic excitations in the bilayer compound. Due to an intrinsic frustration effect involving different canting proclivities of in-plane and out-of-plane neighboring spins, the orbital-mixing terms cannot be gauged away, and play an important role in inducing the anisotropy gap in the spin wave spectrum. 

Since the bilayers are well separated by strontium atoms and the bilayer extent is much smaller than the bilayer repeat distance, the $k_z$ dependence is expected to be small and will be neglected here as in other investigations.\cite{carter2013microscopic,carter2013theory}
For a single bilayer, we consider the tight-binding Hamiltonian in a composite orbital-sublattice-layer basis. The intra-layer terms $\mathcal{H}^{(1)}$ and $\mathcal{H}^{(2)}$ corresponding to the two layers (1) and (2) are as discussed in Section II. In the inter-layer term given below:
\begin{eqnarray}
\mathcal{H}^{(12/21)} 
& = & \sum_{{\bf k} l \sigma s} \psi_{{\bf k} l \sigma s}^{\dagger} 
\begin{pmatrix}
{\epsilon_{\bf k}^{2yz}}^{\prime} & 0 & 0 \\
0  & {\epsilon_{\bf k}^{2xz}}^{\prime}  & 0 \\
0 & 0 & {\epsilon_{\bf k}^{2xy}}^{\prime}  \\
\end{pmatrix} 
\psi_{{\bf k} \bar{l} \sigma s}
+ \sum_{{\bf k} l \sigma s} \psi_{{\bf k} l \sigma s}^{\dagger}
\begin{pmatrix}
t_{4}^{z} & t^{z}_{m} & 0 \\
-t^{z}_{m} & t_{4}^{z} & 0 \\
0 & 0 & t_{1}^{z} \\
\end{pmatrix}
\psi_{{\bf k} \bar{l} \sigma \bar{s}}
\label{three_orb_two_sub_matrix}
\end{eqnarray} 
the two terms correspond to  hopping terms connecting same and opposite sublattices. Here $t_{1}^{z}$ and $t_{4}^{z}$ are the NN inter-layer hopping terms  for the $xy$ and $yz/xz$ orbitals, respectively, while $t^{z}_{m}$ represents the inter-layer orbital mixing between $yz$ and $xz$ orbitals. Arising from the staggered $\rm IrO_6$ octahedral rotations, these anti-symmetric orbital mixing terms have significant magnitude due to the  pronounced 3D character of $yz$ and $xz$ orbitals. The bilayer dispersion terms corresponding to NNN hopping are given by ${\epsilon_{\bf k}^{2yz}}^{\prime} = -2t^z_6\cos k_y$, ${\epsilon_{\bf k}^{2xz}}^{\prime} = -2t^z_6\cos k_x $ and ${\epsilon_{\bf k}^{2xy}}^{\prime}  = -4t^z_2 \cos k_x \cos k_y$ for the three orbitals. 

\begin{figure}
\vspace*{0mm}
\hspace*{0mm}
\psfig{figure=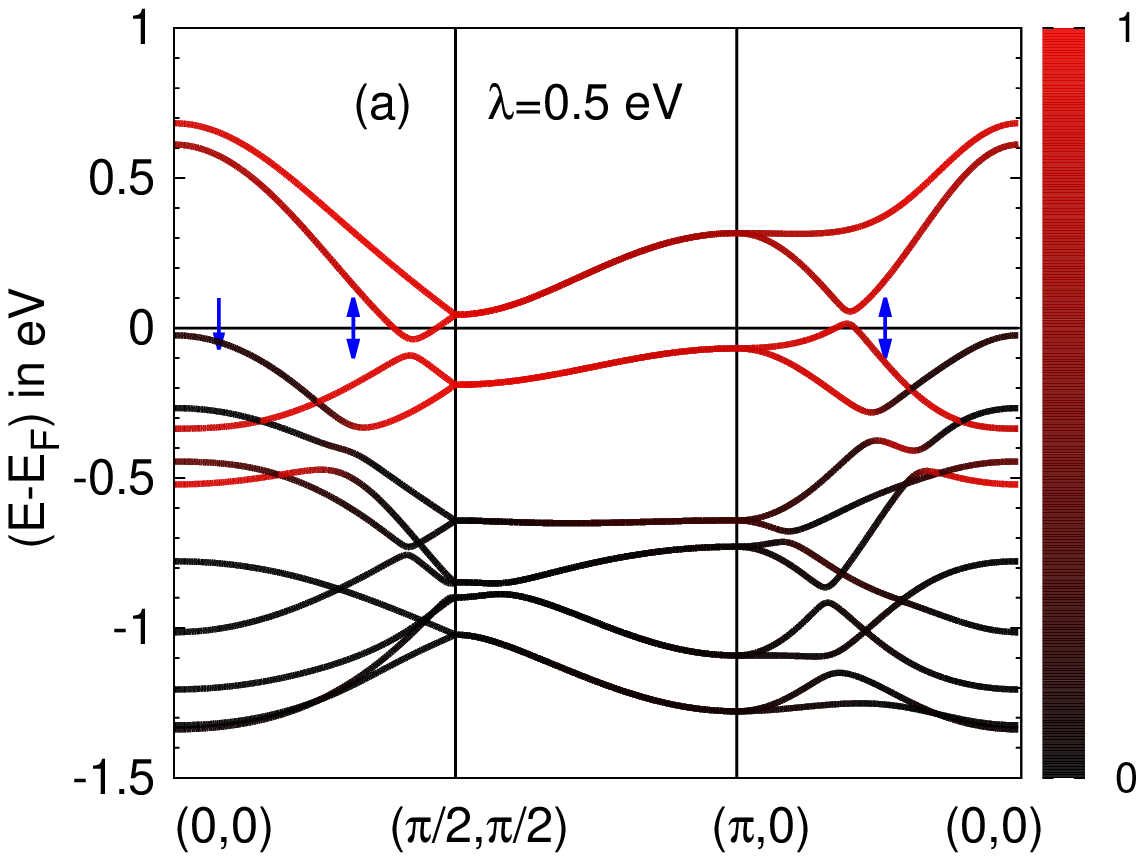,angle=0,width=80mm}
\psfig{figure=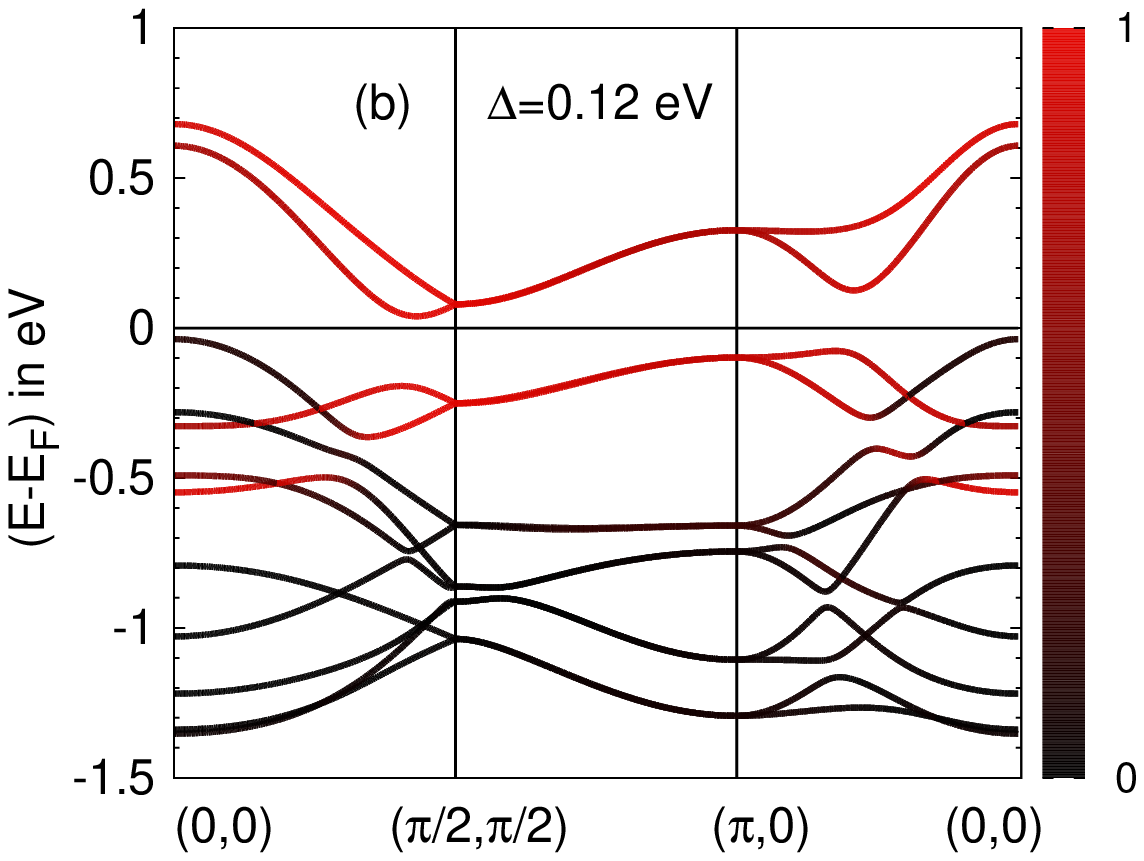,angle=0,width=80mm}
\caption{Electronic band structure for $\rm Sr_3Ir_2O_7$ with SOC only (a), and with both SOC and staggered field (b). The weight of $J=1/2$ (3/2) states are indicated by red (black). The blue arrows (a) show direction of band shifting near Fermi level at larger SOC values.}
\label{bilayer_band}
\end{figure}

We have taken values of the tight binding parameters ($t_1$, $t_2$, $t_3$, $t_4$, $t_5$, $t_{m}$, $t^\prime_{m}$, $t_{1}^{z}$, $t_{4}^{z}$, $t^{z}_{m}$, $t^z_6$, $t^z_2$, $\mu_{xy}$) = (0.2, -0.032, -0.02, 0.188, -0.054, -0.03, 0.022,  0.03, -0.16, 0.072, -0.04, 0, 0) eV as considered earlier.\cite{carter2013microscopic} A small intra-layer third-neighbor hopping ($t_3$) for the $xy$ orbital has been included to incorporate finite zone boundary spin wave dispersion. The ($z$-direction) symmetry-breaking staggered-field term $\mathcal{H}^{(l)}_{\rm sf}$ for both  layers are included as discussed in Sec. II. The calculated band structure of $\rm Sr_3Ir_2O_7$ is shown in Fig. \ref{bilayer_band} at half filling with only SOC (a), and with both SOC and staggered field (b). In units of $t_1$, we have $\lambda = 2.5$ and $\Delta = 0.6$. At intermediate SOC, states near the Fermi level have mostly $J=1/2$ character (red), though there is significant mixing between $J=1/2$ and $3/2$ states (black) below the Fermi level near the point $(0,0)$ in the Brillouin zone.

In comparison to $\rm Sr_2IrO_4$, band structure of $\rm Sr_3Ir_2O_7$ shows a very small Fermi level crossing of the bilayer-split upper $J=1/2$ band near $(\pi/2,\pi/2)$ [Fig. \ref{bilayer_band}(a)]. Band features near the Fermi level strongly depend upon hopping parameters and SOC. With increasing SOC, the low energy states near $(0,0)$ and $(\pi/2,\pi/2)$ shift away from the Fermi level as indicated by blue arrows in Fig.~\ref{bilayer_band}(a). This leads to opening of direct gap across the Fermi level, resulting in band insulating behavior at larger SOC. In the magnetically ordered state [Fig.~\ref{bilayer_band}(b)], the top (bottom) of the occupied (unoccupied) band shifts away from the Fermi level, while maintaining the topology of the non-magnetic bands. At intermediate SOC, the overall band dispersions match reasonably well with experimentally determined bands and first-principles calculations.\cite{wang2013dimensionality, moreschini2014bilayer,moon2008dimensionality,kim2008novel}
 

The addition of an extra $\rm IrO_2$ layer in $\rm Sr_3Ir_2O_7$ dramatically changes both the magnetic ordering and excitation spectrum compared to the single-layer counterpart. The recent RIXS study\cite{boseggia2015evidence} shows two spin wave branches (features B and D) which are well separated in energy and gapped across the entire Brillouin zone. Feature B has sizable dispersion of $85\pm5$ meV and gap of comparable magnitude, whereas feature D is weakly dispersive around $155 \pm 5$ meV and is more clearly visible along the direction $(0,0)$ to $(\pi/2,\pi/2)$. Feature D was not reported in an earlier RIXS study,\cite{kim2012large} where feature B was interpreted as two overlapping spin wave branches (acoustic + optical). In the recent RIXS study,\cite{boseggia2015evidence} the magnetic dispersion is described theoretically in terms of quantum dimer excitations, and features B and D have been interpreted as transverse and longitudinal excitations. 

\begin{figure}
\vspace*{0mm}
\hspace*{0mm}
\psfig{figure=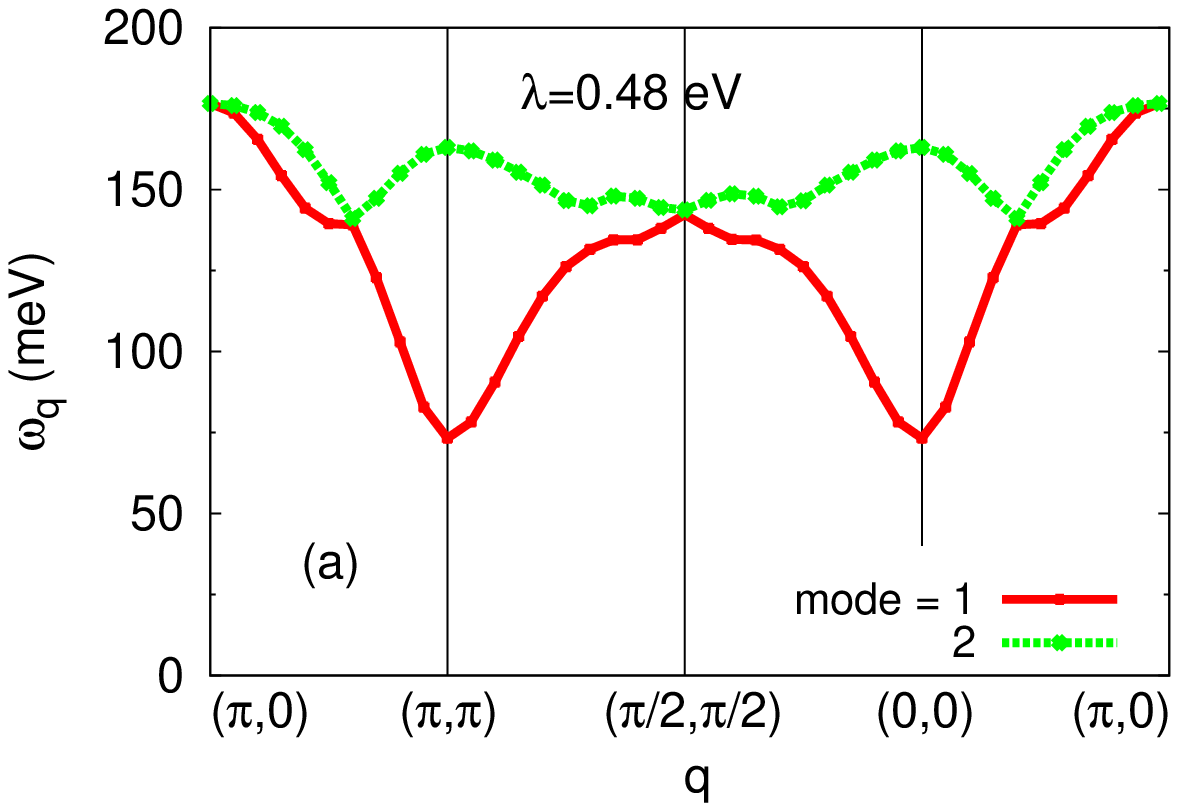,angle=0,width=80mm}
\psfig{figure=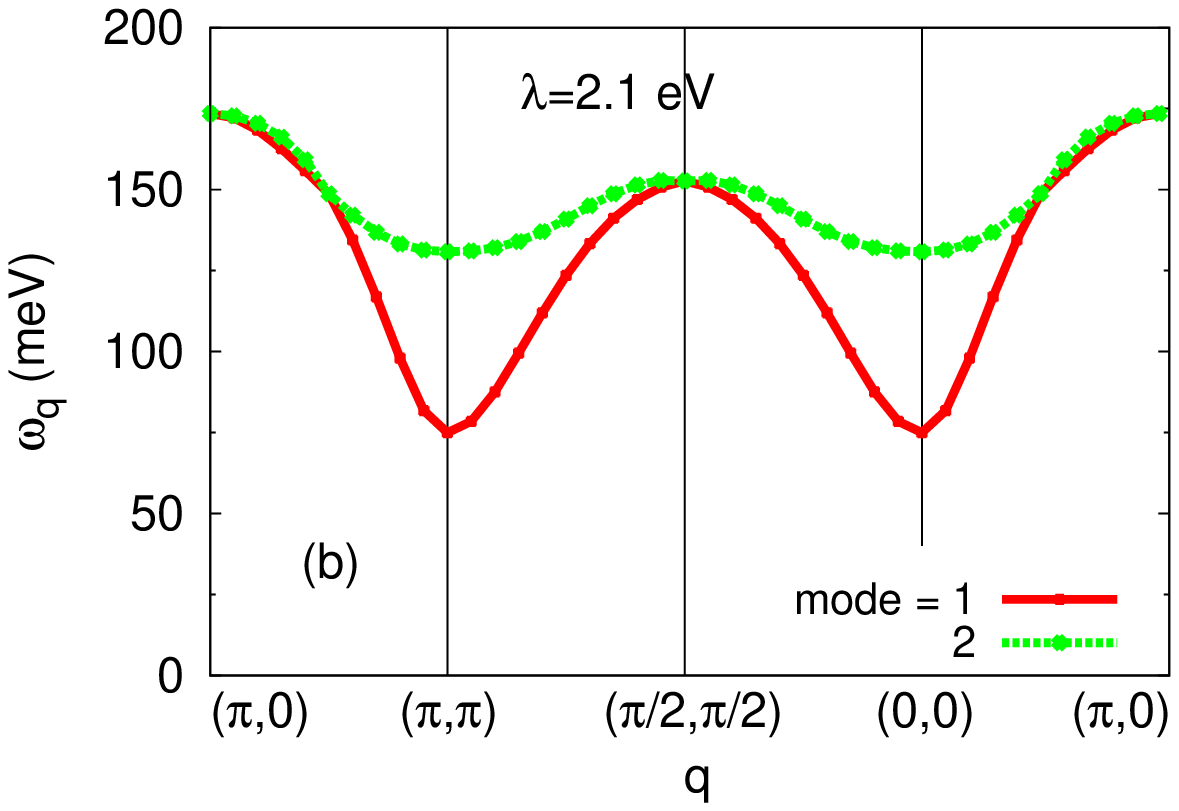,angle=0,width=80mm}
\caption{Calculated spin wave dispersion for $\rm Sr_3Ir_2O_7$ using the three-band-model approach shows (a) dispersion over $\sim$ 100 meV and gap of $\sim$ 75 meV for the first branch. The peak feature of the second branch (a) at ($0,0$) and ($\pi,\pi$) is strongly affected at large SOC (b).}
\label{wq1}
\end{figure}

We have investigated spin waves in $\rm Sr_3Ir_2O_7$ using the three-orbital-model formalism described in the previous section by extending the evaluation of $[\chi^0({\bf q},\omega)]$ in Eq. (8) to the composite two-sublattice, two-layer basis. Fig.~\ref{wq1}(a) shows the calculated spin wave dispersion for $t_1 = 1$ (energy scale unit = 213 meV), $\lambda = 2.25$, and $\Delta = 1$ (corresponding to $U = 3.52$ and $m$ = 0.57). All other tight binding parameter values are as taken earlier for the electronic band structure. The smaller staggered magnetization in comparison to $\rm Sr_2IrO_4$ is due to the inter-layer hopping and enhanced electron itineracy, resulting in a weaker insulator. The values $\lambda \approx 0.48$ eV and $U \approx 0.75$ eV taken above lie within the estimated range for bilayer iridates. \cite{kim2008novel,moreschini2014bilayer} Detailed comparison of the calculated spin wave dispersion with the RIXS spectra is discussed below. 

The calculated spin wave dispersion shows two spin wave modes corresponding to acoustic and optical branches as expected for the bilayer system. The first branch in Fig. \ref{wq1}(a) having dispersion in the $75-175$ meV energy range qualitatively matches with feature B of the RIXS spectra. The spin wave gap depends strongly on the ratios $t^{z}_{m}/t_4^z$ and $t_m/t_1$. We have set $t^{z}_{m}/t_4^z$ $\approx$ 6.5 $t_{m}/t_1$ in our calculation for qualitative match with the RIXS spectrum. This is approximately double of the estimated minimum value required to stabilize the $c$-axis collinear AF ground state over the canted in-plane state.\cite{carter2013microscopic,carter2013theory} The interlayer orbital mixing term $t^{z}_m$ is relatively larger as the neighboring $yz$, $xz$ orbitals are in more enmeshed configuration compared to intra-layer neighbors.

As mentioned earlier, the spin wave gap arises from the frustration effect due to different canting proclivities of in-plane and out-of-plane neighboring spins. This is illustrated in Appendix D in terms of the effective spin model (including the anisotropic spin interaction terms) which follows from the strong-coupling expansion of the single-band Hubbard model with spin-dependent hopping terms. It is the orbital mixing term in the three-band model which generates the spin-dependent hopping terms in the $J=1/2$ sector (Appendix B). The interlayer orbital mixing term in the bilayer compound has appreciable magnitude due to the 3D shape of the spin-orbit entangled states. Therefore, the anisotropic spin couplings and observed spin reorientation with the addition of the $\rm IrO_2$ layer in $\rm Sr_3Ir_2O_7$ are natural consequences of the novel $J=1/2$ electronic ground state. 



Returning now to the second spin wave branch which peaks at $(\pi,\pi)$ and $(0,0)$, where it is most prominent and clearly separated from the first branch. The energy scale and weakly dispersive nature of this branch qualitatively match with the feature D in RIXS spectra, which shows a distinguishable peak near $(0,0)$.\cite{boseggia2015evidence} Although the peak near $(\pi,\pi)$ is not distinguishable in feature D, color map of RIXS spectra near $(\pi,\pi)$ does show substantial spectral weight at higher energies. At large SOC, we find a dip in the second branch at $(\pi,\pi)$ and $(0,0)$ instead of the peak feature [Fig. \ref{wq1}(b)], indicating that the peak feature is connected to mixing between the $J=1/2$ and 3/2 sectors. The spin wave gap is seen to be unaffected by this mixing [Fig. \ref{wq1}(b)]. 

If mixing between $J=1/2$ and 3/2 sectors is neglected in the three-band model, then the effective one-band Hubbard model in the $J=1/2$ sector (Appendix B) shows the same dip feature in the second spin wave branch. This is consistent with the result obtained above in the three-band model at large SOC [Fig. \ref{wq1}(b)], confirming that the peak feature is absent when the $J=1/2$ and 3/2 sectors are decoupled. The opposite sign of the second-neighbor hopping term considered in the effective one-band Hubbard model (Appendix C) reflects the renormalization of hopping parameters in the $J=1/2$ sector due to mixing with the $J=3/2$ sector.

In summary, the main features of our calculated spin wave dispersion are in qualitative agreement with experimental results, with the first and second branches being comparable to features B and D in the RIXS spectra. The peak feature of second spin wave branch near $(0,0)$ and $(\pi,\pi)$ provides evidence of finite mixing between the $J=1/2$ and 3/2 sectors. 

\section{Conclusions}
Magnetic excitations in single-layer and bilayer iridates were studied using a three-orbital-model approach which allows for finite $U$, finite $\lambda$, and finite orbital mixing to be incorporated on equal footing. With realistic three-orbital-model parameters for $\rm Sr_2 Ir O_4$ and $\rm Sr_3 Ir_2 O_7$, the calculated spin wave dispersions are in good agreement with RIXS data, including the important features of strong zone boundary dispersion in $\rm Sr_2 Ir O_4$ and the large anisotropy-induced gap in $\rm Sr_3 Ir_2 O_7$, thus providing a unified understanding of magnetic excitations as well as the electronic band structure. By comparing the calculations for realistic and large SOC values, the RIXS data was shown to provide evidence of mixing between the $J=1/2$ and 3/2 sectors in both iridate compounds. As the extended-range effective magnetic interactions responsible for the strong zone-boundary dispersion, conventionally introduced as phenomenological exchange constants in spin models, arise naturally as finite $U$ effects in the particle-hole exchange process, the itinerant electron approach provides a microscopic description of magnetic excitations in iridates in terms of the spin-orbital entangled states.

\appendix 

\section{AF zone-boundary dispersion of spin waves in the Hubbard model}

Spin waves in the half-filled Hubbard model on a square lattice show no dispersion along the AF zone boundary at large $U$. Only first-neighbor AF spin couplings $J=4t^2/U$ are generated, mixing between opposite sublattices (connected by hopping $t$) is negligible beyond order $t^2$ (super-exchange) level, and the RPA level spin wave dispersion $\omega_{\bf q} = 2J \sqrt {1-\gamma_{\bf q} ^2}$ is flat along the AF zone boundary where $\gamma_{\bf q}=0$. However, at finite $U$, higher-order sublattice mixing terms become important, which generate extended-range spin couplings, resulting in strong zone-boundary dispersion. The salient finite-$U$ features are summarized here in view of the direct relevance to magnetic excitations in Iridates as seen from strong zone-boundary dispersion in the RIXS spin wave data. 

By carrying out the strong-coupling expansion to next-to-leading order (including $t^4/U^4$ terms), spin wave dispersion has been obtained earlier for the $t-t'$ Hubbard model, which accounts for the measured zone boundary dispersion in $\rm La_2 Cu O_4$ from high-resolution neutron scattering studies.\cite{singh1993spin,singh2002spin} Extending the above analysis to the $t-t'-t''-t'''$ Hubbard model including up to fourth-neighbor hopping terms, the spin wave energy is obtained as:
\begin{eqnarray}
\left ( \frac{\omega_{\bf q}}{2J} \right )^2 
& = & (1 - \gamma_{\bf q}^{2}) - \frac{4t^2}{U^2} 
\left (6 + 3 \gamma_{\bf q}^{\prime} - 9 \gamma_{\bf q}^2 \right ) -\frac{{2t^\prime}^2}{t^2}(1-\gamma_{\bf q}^{\prime}) - \frac{2{t^{\prime\prime}}^2}{t^2}(1-\gamma_{\bf q}^{\prime\prime}) \nonumber \\ 
& + & \frac{8{t^{\prime\prime\prime}}^2}{t^2}(1-\gamma_{\bf q}^{\prime}) \gamma_{\bf q}^2
\label{wqexpn}
\end{eqnarray}
where $\gamma_{\bf q} = (\cos{q_x}+\cos{q_y})/2$, $\gamma_{\bf q}^{\prime} = \cos{q_x}\cos{q_y}$, and $\gamma_{\bf q}^{\prime\prime} = (\cos{2q_x}+\cos{2q_y})/2$. It is evident from the above expression that extended-range (2nd/3rd/4th neighbor) effective spin couplings are generated at finite $U$ corresponding to the different ${\bf q}$ dependent terms. In order to quantify spin wave dispersion along the AF zone boundary $(\pi,0) \rightarrow (\pi/2,\pi/2) \rightarrow (0,\pi)$, we consider the spin wave energies:
\begin{eqnarray}
\omega(\pi,0) & = & 2J \left( 1 - \frac{12 t^2}{U^2} - \frac{4 {t^\prime}^2}{t^2}\right)^{1/2} \\ \nonumber
\omega(\pi/2,\pi/2) & = & 2J \left( 1 - \frac{24 t^2}{U^2} - \frac{2 {t^\prime}^2}{t^2}  - \frac{4 {t^{\prime \prime}}^2}{t^2}\right)^{1/2}
\label{wqafzb}
\end{eqnarray}
which indicate that lower $U$ and finite third-neighbor hopping $t''$  increase the dispersion and thus reduce the ratio $\omega(\pi/2,\pi/2)/ \omega(\pi,0)$, whereas $t'$ has opposite effect.

\begin{figure}
\vspace*{0mm}
\hspace*{0mm}
\psfig{figure=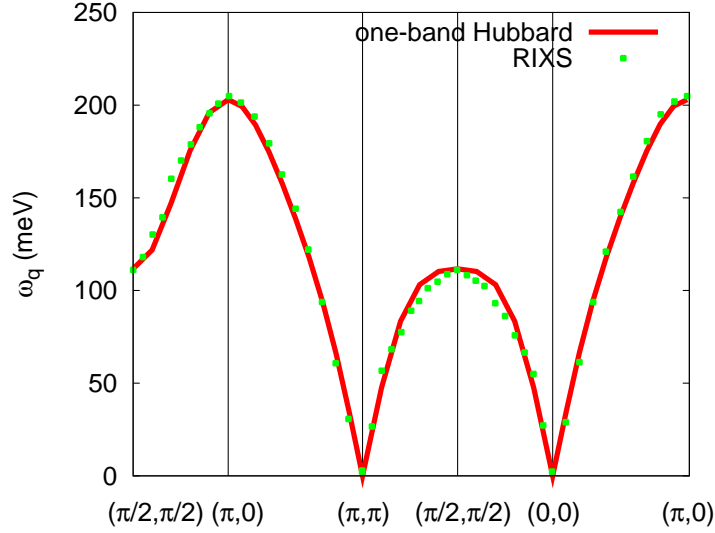,angle=0,width=100mm}
\caption{Calculated spin wave dispersion in the AF state of the one-band Hubbard model along symmetry directions in the Brillouin zone and comparison with RIXS data for $\rm Sr_2 Ir O_4$.}
\label{fig1_app}
\end{figure}

Calculated spin wave dispersion fitted to the RIXS data along symmetry directions in the Brillouin zone is shown in Fig. \ref{fig1_app} for the following parameters: $t=1$ (unit of energy scale $=160$ meV), $t'=0.0$, $t''=0.23$, $t'''=0$, and $\Delta=1.9$ (corresponding to $U=5t=0.8$ eV and $m=0.76$). The spin wave fit to RIXS data shows that magnetic excitations in Iridates are best described, within the  $J=1/2$ sector, by the effective one-band Hubbard model in the intermediate $U$ range. The extended-range (further-neighbor) spin couplings conventionally introduced as phenomenological Heisenberg exchange constants to fit the RIXS data, including the ferromagnetic $J^\prime$ coupling in particular, are seen to arise naturally as finite $U$ effects, thus providing a microscopic understanding of magnetic excitations. 

Mapping from the three-orbital model to the one-band Hubbard model formally involves the following steps. i) The $J=1/2$ sector in Eq. (B1) yields the one-band Hubbard model with spin-dependent hopping term arising from the hopping terms involving orbital mixing between $yz$ and $xz$ orbitals due to staggered octahedral rotations. Mixing with the $J=3/2$ sector is neglected at this level. ii) The spin-dependent hopping term can be gauged away, resulting in a conventional one-band Hubbard model (Appendix C). iii) Strong mixing between the $J=1/2$ and 3/2 sectors at finite SOC will effectively renormalize the $J=1/2$ sector. Neglecting momentum dependence, mixing will effectively reduce the staggered field $\Delta$ (and therefore $U$) by unequally pushing up the two bands in the $J=1/2$ sector ($\approx \delta_{\rm mix}^2 / (\frac{3}{2} \lambda \pm \Delta)$). Momentum-dependent mixing can be incorporated through renormalization of the hopping terms. 

\section{Hamiltonian representation in the $|J,m \rangle$ basis}

Using Eq. (\ref{jmbasis}), matrix representation of the three-orbital model is given below in the $|J,m \rangle$ basis, showing the Hamiltonian matrix elements in the $J=1/2$ and 3/2 sectors as well as the mixing  between the two sectors arising from the hopping terms. As seen below in the $J=1/2$ sector, the hopping term ${\cal E}^{yz|xz}_{\bf k}$ involving orbital mixing between $yz$ and $xz$ orbitals leads to the spin-dependent hopping term which breaks spin-rotation symmetry. 

\begin{equation}
\left [
\begin{array}{cc|cccc}
\lambda & 0 & 0 & 0 & 0 & 0\\
+ \frac{ \epsilon^{yz}_{\bf k} + \epsilon^{xz}_{\bf k} + \epsilon^{xy}_{\bf k} }{3} 
& 
&   
& + \frac{ \epsilon_{\bf k}^{yz} + \epsilon_{\bf k}^{xz} - 2\epsilon_{\bf k}^{xy} }{3\sqrt{2}}
&   
& + \frac{ \epsilon_{\bf k}^{yz} - \epsilon_{\bf k}^{xz}}{\sqrt{6}} \\
+ \frac{2 i \epsilon^{yz|xz}_{\bf k} }{3} 
& & & +\frac{2i\epsilon^{yz|xz}_{\bf k} }{3\sqrt{2}} 
& & 
\\
& \lambda & 0 & 0 & 0 & 0 \\
& + \frac{ \epsilon^{yz}_{\bf k} + \epsilon^{xz}_{\bf k} + \epsilon^{xy}_{\bf k} }{3} 
& + \frac{ \epsilon_{\bf k}^{yz} - \epsilon_{\bf k}^{xz}}{\sqrt{6}}
& 
& + \frac{ \epsilon_{\bf k}^{yz} + \epsilon_{\bf k}^{xz} - 2\epsilon_{\bf k}^{xy} }{3\sqrt{2}}
& \\  
& - \frac{2 i \epsilon^{yz|xz}_{\bf k} }{3}
& & & - \frac{2i \epsilon^{yz|xz}_{\bf k} }{3\sqrt{2}} & 
\\ \hline 
& & -\frac{\lambda}{2} & 0 & 0 & 0 \\
& & + \frac{ \epsilon^{yz}_{\bf k} + \epsilon^{xz}_{\bf k} }{2}
& 
& + \frac{ \epsilon_{\bf k}^{yz} - \epsilon_{\bf k}^{xz}}{2\sqrt{3}}
& \\
& & + \frac{2i\epsilon^{yz|xz}_{\bf k} }{2} & & & 
\\ 
& & & -\frac{\lambda}{2} & 0 & 0 \\
& & & + \frac{ \epsilon^{yz}_{\bf k} + \epsilon^{xz}_{\bf k} + 4\epsilon^{xy}_{\bf k} }{6}
&  
& + \frac{ \epsilon_{\bf k}^{yz} - \epsilon_{\bf k}^{xz}}{2\sqrt{3}} \\
& & & + \frac{2i\epsilon^{yz|xz}_{\bf k} }{6} & & 
\\
& & & & -\frac{\lambda}{2} & 0 \\
& & & & + \frac{ \epsilon^{yz}_{\bf k} + \epsilon^{xz}_{\bf k} + 4\epsilon^{xy}_{\bf k} }{6} 
& \\
& & & & -\frac{2i\epsilon^{yz|xz}_{\bf k} }{6} & 
\\
& & & & & -\frac{\lambda}{2} \\
& & & & & + \frac{ \epsilon^{yz}_{\bf k} + \epsilon^{xz}_{\bf k} }{2} \\
& & & & & -\frac{2i\epsilon^{yz|xz}_{\bf k} }{2}

\end{array}
\right ]
\label{matrix_kin}
\end{equation} 

\section{Gauge transformation}

We consider a single-band Hubbard model with hopping term:
\begin{equation}
\mathcal{H}_0 = -\sum_{\langle ij \rangle} \Psi_i ^\dagger \left [ 
\begin{array}{lr} t+it' & 0 \\ 0 & t-it' \end{array} \right ] \Psi_j
+ {\rm H.c.}
\end{equation}
which breaks spin rotation symmetry due to the spin-dependent hopping term $t^\prime$ originating from the orbital mixing ($yz,xz$) due to staggered octahedral rotation. This Hubbard model describes the $J=1/2$ sector of the three band model with orbital mixing. Strong-coupling expansion yields an effective spin model with PD and DM interactions besides the usual isotropic Heisenberg term (Appendix D), and the classical energy minimum is independent of polar angle $\theta$ if spins are canted at angle $\tan \phi = t'/t$. 

Rewriting the above Hamiltonian as:
\begin{equation}
\mathcal{H}_0 = -\tilde{t} \sum_{\langle ij \rangle} \Psi_i ^\dagger \left [ 
\begin{array}{lr} e^{i \phi} & 0 \\ 0 & e^{-i \phi} \end{array} \right ] \Psi_j + {\rm H.c.}
\end{equation}
where $\tilde{t} = \sqrt{t^2 + t^{'2}}$ and $\tan \phi = t'/t$, the spin-dependent diagonal terms can be absorbed via a spin- and site-dependent gauge transformation:
\begin{equation}
\Psi_j = \left ( \begin{array}{c} a_{j\uparrow} \\ a_{j\downarrow} \end{array} \right ) \rightarrow \left ( \begin{array}{c} e^{i \epsilon_j \phi /2} a_{j\uparrow} \\ e^{-i \epsilon_j \phi /2} a_{j\downarrow} \end{array} \right ) 
\end{equation}
where $\epsilon_j = \pm 1$ for A/B sublattices. The matrix in Eq. (C2) transforms to unit matrix and spin rotation symmetry is now restored. The gauge transformation corresponds to local coordinate-axes rotation so that the originally canted spins are now perfectly antiparallel. 

As the classical energy is invariant under global $\theta$ and $\phi$ variation (provided canting angle is maintained), there exist gapless Goldstone modes. The Goldstone mode eigenvector $(e^{i \phi} \;\;\; -e^{-i \phi})$ takes care of the required canting in the $xy$ plane to ensure zero energy cost. If the 4th neighbor hopping term $t_4$ is included (also connecting opposite sublattices), then corresponding spin-dependent part $t_4 '$ must be included with same ratio $t_4 '/t_4 = t'/t = \tan \phi$ in order for the Goldstone mode to remain gapless. For the bilayer, similar condition $t_z '/t_z = t'/t = \tan \phi$ must be satisfied for the spin wave spectrum to remain gapless. If the ratio $t_z '/t_z \neq t'/t$, then the only way to simultaneously satisfy the two different canting conditions for in-plane and out-of-plane neighbors is $c$-axis alignment, and spin wave spectrum in this case is gapped. Spin wave spectrum will therefore show anisotropy gap if $t_4 '/t_4 \neq t'/t$ (single layer) and $t_z '/t_z \neq t'/t$ (bilayer). 

\begin{figure}
\vspace*{0mm}
\hspace*{0mm}
\psfig{figure=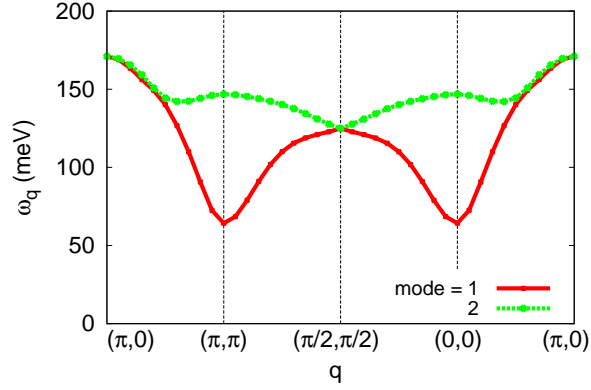,angle=0,width=80mm}
\caption{Calculated spin wave dispersion in the AF state of the bilayer one-band Hubbard model with spin-dependent hopping terms.}
\label{bilayer_one_band}
\end{figure}

The calculated spin wave dispersion for the bilayer Hubbard model at half filling is shown in Fig.~\ref{bilayer_one_band}. Here we have considered the parameters: $t = 1$ (energy scale unit = 150 meV), $t_2 = 0.3$, $t_3 = -0.1$, $t ^\prime = -0.2$, $t_z = -0.8$, $t_z^\prime = 0.55$ and $\Delta = 1.9$, which  correspond to $U = 5.19t = 0.78$ eV, $m = 0.73$, and the ratio $t_z '/ t_z \sim 3.5 t '/t$. The calculated dispersion is in qualitative agreement with the three-band-model calculation presented in section IV as well as with the RIXS data for the bilayer compound. The relatively larger inter-layer spin-dependent hopping term $t_z '$ represents the stronger $yz,xz$ orbital mixing compared to in-plane neighbors. The effective hopping parameters considered above are qualitatively comparable with the three-band model parameters. The peak feature of second spin wave branch is obtained for opposite sign of $t_3$, which confirms the effect of mixing between $J=1/2$ and 3/2 sectors. Also, the spin wave calculation explicitly shows that for $t_z '/t_z = t'/t$ the anisotropy gap vanishes. 

\section{Frustration, magnetic order, and anisotropy gap in the bilayer}

The frustration effect due to different canting proclivities for intra-layer and inter-layer pairs of neighboring spins, which results in the preferred c-axis orientation and the spin wave gap, is briefly illustrated below. Starting with the Hubbard model including a spin-dependent antisymmetric hopping term $t^\prime$ (Appendix C), strong-coupling expansion yields:
\begin{equation}
H_{\rm eff} = \frac{4(t^2 - {t^\prime}^2)}{U} \sum_{\langle ij \rangle} {\bf S}_i . {\bf S}_j + \frac{8{t^\prime}^2}{U} \sum_{\langle ij \rangle} S_i ^z S_j ^z + \frac{8t t^\prime}{U} \sum_{\langle ij \rangle} 
({\bf S}_i \times {\bf S}_j ). \hat{z} - \frac{(t^2 + {t^\prime}^2)}{U} \sum_{\langle ij \rangle} 1 
\end{equation}
which includes the PD and DM anisotropic spin interactions along with the usual isotropic AF interaction for spin quantum number $S=1/2$.  

We consider a NN pair of spins on opposite sublattices, with initial AF orientation along the $z$ direction. If the spin orientations are changed to polar angle $\theta$ and canting angle $\phi$, the change in the classical interaction energy is obtained as:
\begin{eqnarray}
\Delta E (\theta,\phi) &=& 
\left [ \frac{4(t^2 - {t^\prime}^2)}{U} S^2 \sin ^2 \theta (1 - \cos 2 \phi) \right ]_{\rm iso} - \left [ \frac{8tt^\prime}{U} S^2 \sin ^2 \theta \sin 2 \phi \right ]_{\rm DM} \nonumber \\ &+&   
\left [ \frac{8{t^\prime}^2}{U}S^2 (1-\cos ^2 \theta) \right ]_{\rm PD}  
\end{eqnarray}
explicitly showing the contributions of the isotropic, DM, and PD terms. Minimizing the first two terms with respect to $\phi$ yields the optimal canting angle $\tan \phi^* = t^\prime / t$, and expanding the energy change around the minimum, we obtain:
\begin{eqnarray}
\Delta E (\theta,\phi \approx \phi^*) &=& - \frac{8{t^\prime}^2}{U} S^2 \sin ^2 \theta + \frac{8{t^\prime}^2}{U} S^2 (1 - \cos ^2 \theta )+ 
\alpha (\phi - \phi^*)^2 \nonumber \\
&=& \alpha (\phi - \phi^*)^2
\end{eqnarray}
where the coefficient $\alpha = \frac{8(t^2 + {t^\prime}^2)}{U} S^2 \sin ^2 \theta $. For $\phi = \phi^*$, the energy change is independent of the polar angle $\theta$, showing the degeneracy of classical energy with respect to $\theta$ at the optimal canting angle. 

Now, for two pairs of neighboring spins, one intra-layer and another inter-layer, with optimal canting angles given by $\tan \phi_1 ^* = t_1 ^\prime / t_1$ and $\tan \phi_2 ^* = t_2 ^\prime / t_2$, the energy change:
\begin{equation}
\Delta E_1 + \Delta E_2 = \alpha (\phi - {\phi_1}^*)^2 + 
\beta (\phi - {\phi_2}^*)^2
\end{equation}
where the two optimal canting angles are assumed to be close for simplicity so that that the expansion around energy minimum is valid for both pairs of spins. In the un-frustrated case ($\phi_1 ^* = \phi_2 ^*$ or equivalently $t_1 ^\prime / t_1 = t_2 ^\prime / t_2)$, the energy change is still independent of the polar angle $\theta$ provided the spins are canted at the common optimal angle. However, in the frustrated case ($\phi_1 ^* \neq \phi_2 ^*$), the energy change acquires dependence on the polar angle $\theta$ through the coefficients, and is clearly minimum (zero) when the coefficients $\alpha=\beta=0$, which requires that $\theta=0$. This accounts for the preferred c-axis orientation and the finite anisotropy gap in the spin wave spectrum.  

\bibliographystyle{apsrev4-1}

\bibliography{ref.bib}

\end{document}